\documentclass[aps,prc,nofootinbib,groupedauthor,onecolumn,12pt,floatfix]{revtex4-2}
\usepackage{graphicx}
\usepackage{natbib}
\usepackage{xcolor}

\usepackage{subfig}
\usepackage{float}
\usepackage{amsmath}
\usepackage{hyperref}
\usepackage{changes} 
\usepackage{appendix}

\begin{document}

\title{Scattering phase shifts from overlap relations in the $J$-matrix method
}

\author{Calvin W. Johnson}


\author{Bui Minh Loc}

\author{Austin Keller}

\author{Kenneth M. Nollett}

\address{San Diego State University,
5500 Campanile Drive, San Diego, CA 92182-1233}

\begin{abstract}
 We present an approach to the quantum mechanical 
scattering problem in which computation of a wave
function and determination of scattering parameters are separate 
steps.  This approach is built on the $J$-matrix formalism, but in 
place of the usual $J$-matrix procedures, we introduce a novel
formula that extracts phase shifts accurately 
from scattering states computed in a square-integrable basis.  This formula is closely
analogous to existing integral relations or overlap integrals in a  (continuous) position basis.  We then demonstrate the extraction of phase shifts  from wave functions  in single-channel potential scattering.  For a soft but realistic nucleon-nucleon potential, Daejeon16, we obtain good phase shifts with as few as six oscillator basis states, using the Lippmann-Schwinger equation to obtain scattering states at arbitrary energies even in this highly truncated space.
These results constitute a step towards calculations in which scattering observables are extracted from scattering states generated
in a many-body framework with square-integrable basis functions, such as the configuration-interaction shell model.
\end{abstract}

\maketitle

\section{Introduction}

\label{sec:intro_general}

States in the continuum are thorny but important challenges for quantum systems: scattering and reactions with scattering states have been key experimental probes for atomic and subatomic systems going back over a century~\cite{rutherfordscattering}.
While continuum (or scattering) states 
for a few constituents are computationally tractable~\cite{gloeckle1996three,glockle2012quantum,PhysRevC.58.58}, systems with many particles that include the continuum remain at the forefront of problems. 

The landscape of approaches for scattering states 
is too vast to fully summarize here~\cite{johnson2020white}.
Broadly speaking, while for some classes of systems and models the calculation of scattering observables need not proceed through generating an explicit wave function, in many cases there are two computational tasks: solving the Schr\"odinger or Lippmann-Schwinger equation to obtain a wave function, and projection of that wave function into incoming and outgoing waves to obtain scattering properties like phase shifts, mixing parameters, and ultimately cross sections, analyzing powers, etc.  This process is straightforward for scattering of point particles from a central potential.  In general, however, continuum problems rapidly become  more difficult as the systems to be modeled become more complicated.  The desirability in nuclear physics of taking many-body techniques that have been successful in bound systems and extending them to continuum states has long been recognized.  A significant barrier in this endeavor is that the square-integrable basis functions used in the many-body methods are in general not easily matched onto free waves at large separation, especially when computational resources limit the convergence of solutions in the region where matching has to occur.  Often (as in no-core shell model with 
resonating group~\cite{PhysRevLett.101.092501,PhysRevC.79.044606} or with continuum method~\cite{PhysRevC.87.034326}, or hyperspherical methods~\cite{viviani2020fancyregularizer,marcucci2020review}), the usual configuration-interaction basis is supplemented in some way that matches the interaction region directly onto the long-range asymptotic region.

The division of the problem into construction and extraction steps may be illustrated simply by single-channel scattering of two particles through a central potential~\cite{newton2013scattering}. 
Given the nonrelativistic kinetic energy $\hat{T} = - \frac{\hbar^2}{2\mu} \nabla^2$, where $\mu$ is the reduced mass, and a rotationally invariant 
 potential $\hat{V}$ (for simplicity we consider here only central potentials and ignore intrinsic spin), stationary states can be expressed in terms of partial waves, $\frac{u(r)}{r} Y_{\ell,m}(\theta,\phi)$, where $Y_{\ell,m}$ is a spherical harmonic for orbital angular momentum $\ell$ and $z$-component $m$. The radial wave function $u$ is found through the radial 
 Schr\"odinger equation,
\begin{equation}
    \frac{d^2 u(r)}{dr^2} - \left[ \frac{\ell(\ell + 1)}{r^2} + \frac{2\mu}{\hbar^2}  V(r) - k^2 \right] u(r) = 0,
    \label{eq:radial_Schrodinger}
\end{equation}
where $k = \sqrt{2 \mu E}/\hbar$ is the wave number and $E$ is the (positive) energy.
Wherever and whenever the potential vanishes,  we have  free solutions,
\begin{equation}
    \frac{d^2 u_\mathrm{free}(r)}{dr^2} - \left[ \frac{\ell(\ell + 1)}{r^2}  - k^2 \right] u_\mathrm{free}(r) = 0.
    \label{eq:free_Schrodinger}
\end{equation}
There are of course two free solutions, the regular solution, $f(r)$, which vanishes at $r=0$, and the irregular solution, $g(r)$, which does not. ({These 
are $kr \,j_\ell (kr)$ and $kr \,y_\ell(kr)$, where $j_\ell$ and $y_\ell$ are spherical Bessel and Neumann 
functions, respectively, which for $\ell=0$ become $\sin (kr)$ and $-\cos(kr)$}).  Assuming the potential has a finite range $R_\mathrm{pot}$, that is, $V(r) = 0$ 
for $r > R_\mathrm{pot}$, 
the full solution $u$  becomes  free  at $r > R_\mathrm{pot}$: $u \rightarrow A f + B g$,
where $A$ and $B$ are constants. This can be written at large $r$ as $ C \sin (kr + \delta - \ell \pi/2)= C[ \cos(\delta)
\sin (kr  - \ell \pi/2)+ \sin(\delta) \cos (kr  - \ell \pi/2)]$,
where the phase shift $\delta$ describes the change in the asymptotic free solution due to the 
presence of the potential---either a shift in the long-distance nodal structure of the wave or, alternatively, the nonzero $B$ required by the potential for a given $A$.   Multichannel problems (e.g.\ nonspherical or spin-dependent potentials, so that values of $\ell$ need not match for incoming and outgoing waves, or inelastic scattering that leaves the projectile or target in an excited state) are analogous but have separate $A$ and $B$ amplitudes in every channel; the full set of $B$ coefficients is then related to the full set of $A$ coefficients by a matrix (equivalent to the $S$-matrix) rather than a single parameter. In this paper we restrict our development to single-channel problems but point at the end toward multichannel extensions.

We have recapitulated this pedagogy because  its representation in the continuous $r$ coordinate is familiar to most readers. Yet all of this, including the free solutions $f(r)$ and $g(r)$, is representable exactly and sometimes analytically in a square integrable basis.  The formalism for this approach is known as the $J$-matrix method because its only requirement is that the kinetic energy operator take the 
form of a tridiagonal or Jacobi matrix in the basis~\cite{PhysRevA.9.1201,PhysRevA.9.1209,alhaidari2008j}; our goal here is to develop techniques applicable in the $J$-matrix formalism that will be useful for nuclear many-body scattering. 

In the matching method in position space, one integrates Eq.~(\ref{eq:radial_Schrodinger}) out to some $r > R_\mathrm{pot}$, and then matches the solution $u$ to $A f + B g$ there.  This could be done by first specifying an energy, or else by imposing a boundary condition at large radius and then finding the function $u$ at one or more of the resulting eigenenergies. Matching has direct analogs in the $J$-matrix formalism~\cite{PhysRevC.94.064320}.  However, the use of explicit matching at large $r$ in general requires an accurate representation of the wave function there, which is often not tractable in many-body calculations because of exponential fall-off of the basis functions~\cite{PhysRevLett.87.063201}.
 
Alternately,  given a wave function, one can compute the phase shift or $S$-matrix through an integral relation involving the scattering potential~\cite{berggren1965overlap,pinkston1965form,rodberg1967introduction,timofeyuk1998one,suzuki2009phase,PhysRevC.81.034002,PhysRevC.86.044330,PhysRevC.108.034001}. Because the scattering potential in typical nuclear problems falls off  rapidly at large distances, this approach is insensitive to errors in the asymptotes of the 
wave function, and in fact works well even for approximate wave functions~\cite{PhysRevC.108.034001}. While this approach has been successful, to date it has been applied  to coordinate-space-based methodologies, namely effective one-body scattering~\cite{suzuki2009phase}, variational and Green's function Monte Carlo calculations~\cite{PhysRevC.86.044330}, and hyperspherical harmonic calculations~\cite{PhysRevC.81.034002}; even in applications where the underlying many-body calculation is the configuration-interaction method, the phase shift was computed via a coordinate-space integral~\cite{timofeyuk1998one}.

Not all many-body methods operate in coordinate space, however. 
Many approaches for bound systems, such as configuration interaction or coupled clusters, use square-integrable bases. In nuclear physics, the most common, though by no means the only, single-particle basis is constructed from the isotropic harmonic oscillator~\cite{lawson1980theory}, also sometimes called the Gaussian basis (not to be confused with a basis constructed from linear combinations of Gaussians as used in atomic and molecular physics~\cite{dunning1977gaussian}). 

In this paper we combine these approaches to derive an analog of the integral relation for phase shifts that is formulated explicitly in an orthonormal basis of square-integrable functions like that used in $J$-matrix calculations, rather than in coordinate space.
This result is new, and our derivation sheds light on the structure and origin of coordinate-space integral relations as well.  To illustrate the method, we focus on the problem of extracting scattering parameters from already-computed scattering states, specifically the phase shift in a single-channel problem.
Our ultimate goal, however, is to prepare for applications in settings where the scattering state is computed using square-integrable basis functions, such as 
configuration-interaction calculations in a harmonic-oscillator shell model basis. Thus, this paper should be seen as an intermediate but important first step.

 We begin by reviewing the $J$-matrix formalism in Sec.~\ref{sec:intro_Jmatrix}.
In Sec.~\ref{continuous} we review the derivation of integral relations for the phase shift in coordinate space. Although it is not the focus of this paper, we illustrate in Fig.~\ref{fig:accuracy} how overlap methods can more accurately extract phase shifts than the standard matching approach.  Then in Sec.~\ref{discrete}, we derive the analog relations, which we call scattering overlap relations, in $L^2$-integrable bases, such as in the $J$-matrix method. Sec.~\ref{explain_g} discusses  details of handling the irregular free solution in a basis of functions regular at the origin.
As a demonstration, we consider in Sec.~\ref{example} examples of simple potential scattering. In the Appendix, we consider issues with respect to highly truncated spaces.

\section{Scattering in $J$-matrix formalisms}

\label{sec:intro_Jmatrix}

As the $J$-matrix formalism~\cite{PhysRevA.9.1201,PhysRevA.9.1209,yamani1975j,PhysRevA.55.265,bang2000p} is not widely known, we briefly review it here. Instead of working in coordinate or momentum space, one expands the system in an orthonormal, $L^2$-integrable basis, with 
angular momentum as a good quantum number. (Again we only consider orbital angular momentum $l$ and do not worry about intrinsic spin for this review.)  
While this might seem contrary to any description of scattering states, the key insight 
is that in some bases, 
the matrix representation of the kinetic energy $\hat{T}$ is tridiagonal, that is, a Jacobi matrix; this in turn allows one to find full and for some bases analytic representations of the regular and irregular free scattering solutions~\cite{PhysRevA.9.1201}. 
(Although we do not discuss it here, in the Laguerre basis this can be extended to including charged-particle scattering~\cite{yamani1975j,bang2000p}. Furthermore, generalizations to any square-integrable basis have also been found~\cite{PhysRevA.64.042703}).  While early applications were applied to atomic systems~\cite{PhysRevA.9.1209,PhysRevA.12.1222,PhysRevA.14.2159}, in recent years the $J$-matrix has been applied to many-body nuclear systems as well~\cite{PhysRevC.94.064320,revai1985note,PhysRevC.63.034606,lurie2004loosely,PhysRevC.70.044005,broeckhove20075h,PhysRevC.79.014610,mazur2017description,PhysRevC.98.044624,mazur2019description,shirokov2019description,PhysRevC.106.064320}.

Our (radial) basis states are  $\phi_{n\ell}(r)$ or, in Dirac notation, $| \phi_{n\ell} \rangle$; as 
we assume orbital angular momentum $\ell$ (and total angular momentum $j$) to be fixed,  we mostly suppress it, leaving only 
the radial nodal quantum number $n = 0, 1, 2, \ldots$. 
Importantly, we assume the basis states are by construction regular at the origin, that is, $\phi_{n\ell}(0)=0.$ We expand a  state in this basis,
\begin{equation}
| u \rangle = \sum_{n=0}^{\infty} u_n | \phi_n\rangle.    
\label{eq:wavevector}
\end{equation}
To find the $u_n$, we
compute the matrix elements of the kinetic energy operator $\hat{T}= \hat{p}^2/2\mu$, where $\mu$ is the (reduced) mass of the scattered particle, and of the potential: 
\begin{equation}
    T_{n,n^\prime} = \langle \phi_n | \hat{T} | \phi_{n^\prime} \rangle , \,\,\,\, 
        V_{n,n^\prime} = \langle \phi_n | \hat{V} | \phi_{n^\prime} \rangle.
\end{equation}
%
%
The amplitudes $u_n$ of scattering states are then solutions of the eigenvalue equation,
\begin{equation}
    \sum_{n^\prime =0}^{\infty} ( T_{n,n^\prime} + V_{n, n^\prime} ) u_{n^\prime} 
    = E u_n, \label{eq:horse}
\end{equation}
where $E = \hbar^2 k^2 / 2 \mu >0$ is the energy of the scattering state, and $k$ is the wave number.

We can also expand free ($\hat{V}=0$) scattering solutions.
Because the basis functions themselves are regular, the regular free solution is straightforward:
\begin{equation}
 f(r) = \sum_{n=0}^\infty f_n \phi_n(r)   
\end{equation}
where $f_n = \langle f | \phi_n \rangle $ is just the projection of $f(r)$ into the chosen basis. 

The $g_n$ are more subtle, because $g(r)$ is irregular at $r=0$, that is, it
has a nonzero value; but 
in coordinate space the basis functions, $\phi_n(r) = \langle r | \phi_n \rangle$ 
are regular and vanish at $r=0$. 
In the standard $J$-matrix approach, at least, 
there is a solution to the conundrum of representing the irregular 
free function. One introduces 
 a function $\tilde{g}(r),$
\begin{equation}
    \tilde{g}(r) =  \sum_{n=0}^\infty g_n \phi_n(r) \label{gtilde}
\end{equation}
that has the property~\cite{PhysRevC.94.064320,yamani1975j,bang2000p,PhysRevC.100.034321}:
\begin{equation}
    \lim_{r \rightarrow \infty} \tilde{g}(r) \rightarrow g(r). \label{limgtilde}
\end{equation}
In Sec.~\ref{explain_g} we discuss for the interested reader how $\tilde{g}(r)$ is the solution to an inhomogeneous 
differential equation. 

For now, however, we focus on this key idea:
in the standard $J$-matrix approach, $T_{n,n^\prime}$ is tridiagonal,  as a property of the basis. (Note that in coordinate space, the simplest discretization also leads to a tridiagonal representation of the kinetic energy.) 
The tridiagonal form for the kinetic energy allows one to express the free  solutions through simple recursion relations. In some bases, such as 
the harmonic oscillator basis, the coefficients of the free solutions are known analytically; we give such an example in Sec.~\ref{example}.


In coordinate space, one assumes that the potential $V(r)$ effectively vanishes for $r > R_\mathrm{pot}$, where $R_\mathrm{pot}$ is some range
for the potential. Furthermore, one typically only integrates 
solutions out to some $r=R_\mathrm{max} >R_\mathrm{pot}$.
The $J$-matrix analog is to assume the potential $\hat{V}$ can be well approximated 
by restricting $V_{n,n^\prime}$ to $n,n^\prime \leq N_\mathrm{pot}$. 
Furthermore, one  introduces  an $N_\mathrm{max} \geq N_\mathrm{pot}$, 
where $N_\mathrm{max}$ replaces infinity as the upper limit in 
Eqs.~(\ref{eq:wavevector}) and (\ref{eq:horse}). Then 
the Hamiltonian matrix $H$  schematically looks like (for $N_\mathrm{pot} = 2$ and $N_\mathrm{max} = 10$):
\begin{equation}
H =
\begin{pmatrix}
T_{00} + V_{00} & T_{01} + V_{01} & T_{02} + V_{02} & 0& \cdots & 0 & 0 \\
T_{10} + V_{10} & T_{11} + V_{11} & T_{12} + V_{12} & 0 & \cdots & 0 & 0 \\
T_{20} + V_{20} & T_{21} + V_{21} & T_{22} + V_{22} & T_{23} & \cdots & 0 & 0 \\
0 & 0 & T_{32} & T_{33} & \cdots & 0 & 0 \\
\vdots & \vdots & \vdots & \vdots & \ddots & \vdots & \vdots \\
0 & 0 & 0 & 0 & \cdots & T_{9 \,9} & T_{9\, 10} \\
0 & 0 & 0 & 0 & \cdots & T_{10\, 9} & T_{10\, 10} \\
\end{pmatrix}
\label{Jmatrix}
\end{equation}

 Eq.~(\ref{Jmatrix}),  a standard assumption in $J$-matrix  theory,
brings up several points.
Because of the assumption that $V_{n,n^\prime} = 0$ for $n$ or $n^\prime > N_\mathrm{pot}$, we can formally write in terms of the phase shift $\delta$, 
\begin{equation}
    u_n = C\left[  \cos (\delta) \, f_n + \sin (\delta) \, g_n \right], \,\,\,\, n > N_\mathrm{pot}.
    \label{scatt_ho1}
\end{equation}
This is of course analogous to the coordinate space solution. 
Here we want to emphasize that in the $J$-matrix 
formalism the kinetic energy and its solutions are represented 
exactly. 
(In fact, it is the potential energy that is approximated: we know of no rigorous
theorem that the matrix elements $V_{n,n^\prime} $ \textit{must}  vanish as 
$n$ gets large; nonetheless, experience shows it works as a practical approximation~\cite{lashko2019properties}.
Furthermore, for $N_\mathrm{pot}$ small and $N_\mathrm{pot} \ll N_\mathrm{max}$, the problem is no 
longer variational. We get around this problem by choosing $N_\mathrm{max}=N_\mathrm{pot}$. For further discussion, 
see the Appendix.)

Even when calculation stops at finite dimension $N_\mathrm{max}$ in the basis, the kinetic energy operator is represented exactly: the $J$-matrix formalism allows calculations of finite dimension $N_\mathrm{max}$ to match onto free waves.
One can, for example, convert Eq.~(\ref{eq:horse}) into a discretized continuum problem by 
imposing a boundary condition such as the ``nodal'' condition $u_{N\mathrm{max} } = 0$ and diagonalizing Eq.~(\ref{Jmatrix}).
Then, quite simply, when an eigenvector $u_n$ satisfying this boundary condition has energy $E$, the matching method gives~\cite{PhysRevC.94.064320}
\begin{equation}\label{tandeltabc}
\tan \delta(E) = - \frac{f_{N_\mathrm{max}+1} }{g_{N_\mathrm{max}+1}}.
\end{equation}
The use of a boundary condition to discretize 
the continuum is only an illustrative example, with the 
disadvantage of giving phase shifts only at the eigenenergies constrained by some value of $N_\mathrm{max}$. In our work below, we use the Lippmann-Schwinger equation to obtain 
 solutions at arbitrary scattering energy.
One can also find the $S$-matrix or phase shift at arbitrary scattering energy from 
the $J$-matrix representation of the Green's function~\cite{PhysRevC.94.064320,bang2000p}.   Our main concern here is to outline a method to extract scattering parameters accurately  from a given state that has already been computed.

\section{Scattering overlap relation}

An alternative to matching at large $r$ or $n$ has been long known, but not widely used. 
Many monographs on scattering, e.g.~\cite{newton2013scattering}  contain 
relevant formulas. 
Because such formulas can be derived from Green's functions, this is sometimes referred to as 
the Green's function formalism~\cite{suzuki2009phase}; it is sometimes carried out in the context of the Kohn variational principle, where leading errors on the resulting scattering matrix are rigorously second order in the wave function errors~\cite{PhysRevC.81.034002}. (As we will show below, 
it is perhaps better understood as a consequence of Green's \textit{theorem}, relating an integral over a region to behavior 
at the boundary of the region, where the `boundary' is now at large distance.) 
In applications to many-body systems, they are sometimes simply referred to as `overlap integrals'~\cite{berggren1965overlap,pinkston1965form,timofeyuk1998one}.
{In this paper we derive the analog relations in both continuous and discrete bases in Sections \ref{continuous} and \ref{discrete}, respectively.}
Because we do not always have explicit integrals, but in all cases wave function overlaps onto specific asymptotic waves are sought, we suggest an alternate, more general terminology: scattering overlap relations (SOR).

\subsection{Scattering overlap relation using continuous position basis}
\label{continuous}

As an illustration, we rederive the SOR used in coordinate space, although our main focus is the application to the $J$-matrix formalism below.
We start with the radial Schr\"odinger equation with a central 
potential $V(r)$, Eq.~(\ref{eq:radial_Schrodinger}).
We continue to assume the central potential $V$ has a finite range, and 
vanishes beyond some radius $R_\mathrm{pot}$. (Although both integral relations~\cite{PhysRevC.81.034002} and the $J$-matrix method~\cite{yamani1975j,bang2000p} can be extended to include the long-range Coulomb potential, we leave that to future work.) 

Because $u(r) = C \sin (kr + \delta - \ell \pi /2)$ for $r \geq R_\mathrm{pot}$,
one can solve for $\delta$ by matching  $u(r)$ onto this form at
large $r$, i.e. following the ``matching'' method as described in Sec.~\ref{sec:intro_general} above. Obtaining an accurate phase shift in this way requires accurate knowledge of $u$ at large $r$. For simple potential scattering, this can be done easily, for example using the Numerov method  to obtain an accurate solution all the way to large $r$ and then matching its logarithmic derivative to the known functional form~\cite{Numerov1924, Numerov1927}, but for many-body systems it is much more challenging.  In Fig.~\ref{fig:accuracy} below, we illustrate how an overlap relation can dramatically improve the accuracy of a phase shift using a low-accuracy calculation of a wave function. 

To declutter our expressions, we absorb various constants in Eq.~(\ref{eq:radial_Schrodinger}),(\ref{eq:free_Schrodinger}) by defining $\mathcal{V} = 2\mu/\hbar^2 V$. 
The first step is to multiply Eq.~\eqref{eq:radial_Schrodinger} by $u_\mathrm{free}$ and Eq.~(\ref{eq:free_Schrodinger}) by $u$, then integrate the difference to obtain
\begin{eqnarray}
    \int_0^\infty \left (u_\mathrm{free} \frac{d^2 u}{dr^2} - u \frac{d^2 u_\mathrm{free}}{dr^2} \right )\, dr & = &\int_0^\infty u_\mathrm{free}(r) \mathcal{V}(r) u(r) \, dr  \label{Green1} \\
   &\equiv&  \langle u_\mathrm{free}|\mathcal{V}|u \rangle. \nonumber
\end{eqnarray}
Integrating the left-hand side, 
\begin{equation}
    \left (u_\mathrm{free} \, \frac{du}{dr} - u \, \frac{d u_\mathrm{free}}{dr} \right )\Big|_{r = 0}^{r \rightarrow \infty} = \langle u_\mathrm{free}|\mathcal{V}|u \rangle. \label{Green2}
\end{equation}
This is an example of a \textit{Green's theorem} as used in Ref.~\cite{Lane1958}.
%
%
Because $\lim_{r\rightarrow \infty} u \rightarrow A f + B g$ and $\tan \delta = -B/A$, we must
extract $B$ and then $A$. 
First let $u_\mathrm{free} = f$. As  both $f = u = 0$ at the origin, then $( f u' -u f^\prime)_{r = 0} = 0 $, 
where $u^\prime = du/dr$,
while at large $r$,
\begin{equation}
    (f u'- u f^\prime)_{r \rightarrow \infty}
    = B (fg^\prime - g f^\prime).
\end{equation}
One  recognizes  $(fg^\prime - gf^\prime)$ as the Wronskian from the theory of second-order differential equations; its value is both independent of $r$ and, as we will show, cancels out. Taking the limits 
in Eq.~\eqref{Green2}  one simply gets
\begin{equation}
    B (fg^\prime - g f^\prime) = \langle f|\mathcal{V}|u \rangle. \label{integralphaseshift}
\end{equation}

Next, repeating the above process but letting $u_\mathrm{free}$ be the irregular free solution $g$:
\begin{equation}
    (g u^\prime  - u g^\prime)\Big|_{r = 0}^{r \rightarrow \infty} = \langle g|\mathcal{V}|u \rangle. \label{eq:surface_terms}
\end{equation}
At large distance,
\begin{equation}
    (g u^\prime  - u g^\prime)_{r \rightarrow \infty} = -A (  f g^\prime -g f^\prime ). \label{eq:wronskian_at_inf}
\end{equation}
{Handling the $r\rightarrow 0$ limit of $(g u^\prime  - u g^\prime)$ in Eq.~(\ref{eq:surface_terms}) requires careful, technical considerations, which we leave to Appendix~\ref{sec:tech_origin}.}
{Here we simply write that limit as $\left (g(0) u'(0) - u(0) g^\prime(0) \right )$ and express Eq.~(\ref{eq:surface_terms}) as }
\begin{equation}
-A (f'g - fg') - \left (g(0) u'(0) - u(0) g^\prime(0) \right )  = \langle g|\mathcal{V}|u \rangle. \label{cosdelta}
\end{equation}
Consequently, between Eq.~(\ref{integralphaseshift}) and (\ref{cosdelta}) one can extract the phase shift, with the Wronskian dividing out:
\begin{eqnarray}
    \tan \delta = \frac{B}{A} & = & - \frac{ \langle f|\mathcal{V}|u \rangle}{g(0) u^\prime(0) - u(0) g^\prime(0) + \langle g|\mathcal{V}|u \rangle} \label{deltacont} \\
\nonumber 
    & = & - \frac{\int_0^\infty f(r) {\cal V}(r) u(r) \, dr  }{g(0) u^\prime(0) - u(0) g^\prime(0) + \int_0^\infty g(r) {\cal V}(r) u(r) \, dr},
\end{eqnarray} 
where we have written out the integral explicitly. Note that in practice 
the integrals are restricted to the range of the potential ${\cal V}$. 
{Terms similar to $g(0) u^\prime(0) - u(0) g^\prime(0)$
have been previously found, e.g., Eq. (7) in \cite{PhysRevC.81.034002};
our analysis, however, is exact and appears to be the most general, applicable for all angular 
momentum $l$, while also illuminating the success of schemes for 
regularizing $g(r)$~\cite{PhysRevC.81.034002,PhysRevC.108.034001,viviani2020fancyregularizer}.

It is also important to note that many monographs on scattering~\cite{newton2013scattering}
relate the scattering amplitude to the numerator
of Eq.~(\ref{deltacont}), that is, $- \langle f|\mathcal{V}|u \rangle$.
This approach, however, implicitly assumes a specific asymptotic normalization  
of the wave functions. Such a normalization can be expensive to enforce, especially in many-body calculations. The usefulness of SOR such as Eq.~(\ref{deltacont}) 
is that any overall normalization is canceled out, making it useful for 
numerical approaches.
}

While other methods for extracting the phase shift require one to evaluate the wave function at $r > R_\mathrm{pot}$, the SOR (also referred to as an integral relation or other names) 
relies upon the wave function within the finite range $R_\mathrm{pot}$ of the 
scattering potential. As such, it is less sensitive to the asymptotic behavior of  an approximately computed scattering state~\cite{PhysRevC.108.034001}.

\begin{figure}[h]
    \includegraphics[width=0.9\textwidth]{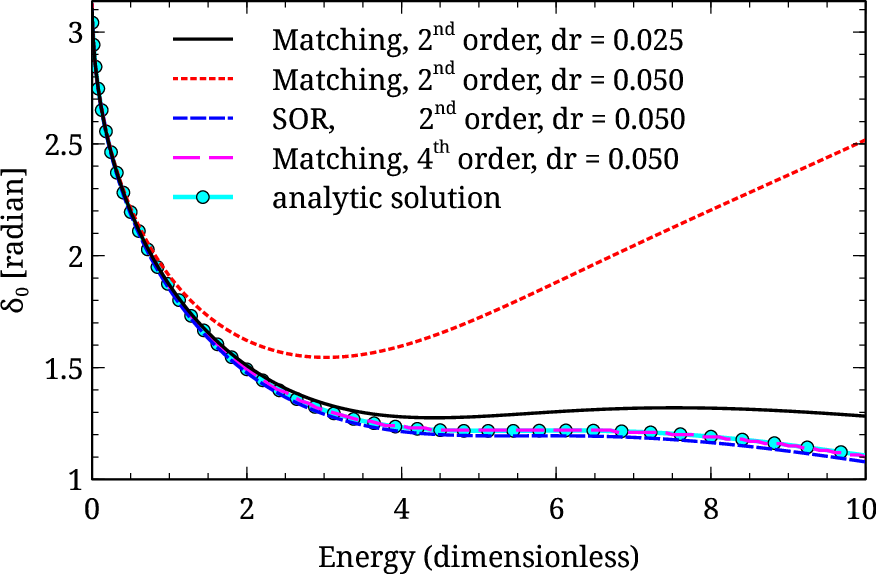}
    \caption{Demonstration of the improved accuracy using scattering 
    overlap relations (SOR), for the $s$-wave phase shifts 
    for a square well with dimensionless radius $R=1$ and depth $V_0 = -5$; in addition, $\hbar=\mu=1$.
    We compare: the analytic phase shifts (line with open circles); 
    phase shifts from matching solutions using Numerov, a method with 
    fourth-order accuracy (long dashed line) that agrees nearly exactly with the analytic solutions; phase shifts from diagonalizing a coordinate-space discretization and matching via Eq.~(\ref{tandeltabc}), a method with second-order accuracy, using dimensionless lattice spacing of 0.025 (solid line) and 0.050 (dotted line); and, finally, phase shifts from the SOR using the same second-order solution with spacing of 0.050 (short dashed line).  The SOR phase shifts (short dashed line) are much more accurate than the very poor phase shifts from matching using the exact same second-order solution (dotted line).
 }
    \label{fig:accuracy}
\end{figure}

{ The robustness of the SOR is demonstrated in Fig.~\ref{fig:accuracy}. We generated solutions $u(r)$ of Eq.~(\ref{eq:radial_Schrodinger}), 
with a square well potential from two different methods: by 
integrating using the fourth-order Numerov method, with 
a (dimensionless) coordinate lattice spacing $dr = 0.050$, and
by diagonalizing the discretized coordinate space Hamiltonian, 
which because of its second-order approximation to the kinetic energy,
is a second-order method; this latter we solved using lattice spacings 
of $dr = 0.050$ and $0.025$.
 When the phase shifts were found 
by matching to the asymptotic solution at large $r$, the fourth-order solution yielded extremely good agreement with the analytic phase shifts, but the second-order approximation performed poorly, especially with the larger lattice spacing. When used in Eq.~(\ref{deltacont}), however, the `worse case' 
solution---the second-order solution using the larger spacing---nonetheless
 yielded significantly more accurate phase shifts, nearly as accurate as the fourth-order matching results.  In one dimension, of course, it is easy to use an accurate, high-order 
method, but this simple exercise illustrates the utility of a SOR, 
especially for cases such as many-body systems where high accuracy comes 
with a high cost.}
{In principle one could apply SOR to wave functions calculated by any method, 
such as the well-known portfolio of R-matrix calculations~\cite{descouvemont2010}, which also carry out matching (explicitly or implicitly) via either a fixed logarithmic derivative or a Bloch operator. Application of SOR to that method’s wave functions should provide results comparable with or better than explicit matching.
}

In the next section, we derive an analogous relation but for square-integrable bases, such as harmonic oscillator or Laguerre bases.

\subsection{Scattering overlap relation using a square-integrable basis}

\label{discrete}

Now we derive the SOR on a discrete basis of square-integrable orthogonal functions 
 suitable for the $J$-matrix methods  of in Sec.~\ref{sec:intro_Jmatrix}, instead of in the continuous coordinate-space basis.

We start with Eq.~(\ref{eq:wavevector}) as the full scattering solution, with coefficients $u_n$  satisfying  Eq.~(\ref{eq:horse}). 
We follow the standard $J$-matrix assumption that the matrix elements  of the kinetic energy $T_{n,n^\prime}$ are tridiagonal, which means that 
at  large $n > N_\mathrm{pot}$, beyond the ``range'' of the truncated representation of the potential, the $u_n$ satisfies the (free) tridiagonal
recursion relation
\begin{equation}
    T_{n,n-1}u_{n-1} + (T_{n,n}-E) u_n + T_{n,n+1} u_{n+1} = 0, \,\,\,  n > N_\mathrm{pot}+1
    \label{eq:tridiagonal_KE}
\end{equation}
Such tridiagonal recursion relations have two independent solutions,
$f_n$ and $g_n$. Here $f_n$ corresponds to the `regular' solution, and 
satisfies at $n=0$,
\begin{equation}
    (T_{0,0}-E) f_0 + T_{0,1}f_1 = 0. \label{eq:regular_Jmatrix}
\end{equation}
Remember that here the index $n=0$ denotes not the origin but the radial basis 
function with zero radial nodes.
One can show that $f_n = \langle f | \phi_n\rangle$ ~\cite{yamani1975j}, so that the 
identification with the regular solution is obvious. 
The second, linearly independent solution $g_n$ 
\textit{cannot} satisfy Eq.~(\ref{eq:regular_Jmatrix}), or else 
$g_n$ would be proportional to $f_n$.  Hence it must satisfy the 
discrete equivalent of a $\delta$-function  inhomogeneity
(so that the inhomogeneous term is now a Kronecker-$\delta$ in function space rather than a Dirac-$\delta$ in coordinate space),
\begin{equation}
    (T_{0,0}-E) g_0 + T_{0,1}g_1 = \alpha_0.  \label{eq:irregular_Jmatrix}
\end{equation}
Note that $\alpha_0$ is not an input parameter but a quantity computed 
using Eq.~(\ref{eq:irregular_Jmatrix}) and the known irregular coefficients $g_0, g_1$. It is related to the inhomogeneous 
differential equation \cite{yamani1975j} for which $\tilde{g}(r) $ is the solution, 
and in Sec.~\ref{explain_g} we find the value of $\alpha_0$ in terms of the Wronskian of $f$ and $g$.  Thus, at large $n > N_\mathrm{pot}$, we must have 
\begin{equation}
    u_n = A f_n + B g_n, \,\,\,\, n > N_\mathrm{pot}.
    \label{eq:l2asympote}
\end{equation}

Now we find the $J$-matrix analog of Eq.~(\ref{Green1}). To do that, we make explicit use of the tridiagonal matrix representation of the kinetic energy as in Eq.~(\ref{eq:tridiagonal_KE}), but including the potential energy~\cite{bang2000p}, which truncates at $N_\mathrm{pot}$
\begin{equation}
    T_{n,n-1}u_{n-1} + (T_{n,n} - E)u_n + T_{n,n+1}u_{n+1} = -\sum_{m=0}^{N_\mathrm{pot}} V_{n,m}u_m, \label{recursion_scat}
\end{equation}
(note for $n > N_\mathrm{pot}$, the right-hand side is zero), while for the regular free solution,
\begin{equation}
    T_{n,n-1}f_{n-1} + (T_{n,n} - E)f_n + T_{n,n+1}f_{n+1} = 0. \label{recursion_free}
\end{equation}
Both Eq.~(\ref{recursion_scat}) and (\ref{recursion_free}) assume $n > 0$. For $n=0$ 
we have 
\begin{eqnarray}
    (T_{0,0} - E) u_0 + T_{0,1}u_1 &=& - \sum_{m = 0}^{N_\mathrm{pot}} V_{0, m} u_m, \label{recursion_scat0}\\
    (T_{0,0} - E) f_0 + T_{0,1}f_1 &=& 0 \label{recursion_free0}.
\end{eqnarray}

Now multiply Eq.~\eqref{recursion_scat} by $f_n$ and Eq.~\eqref{recursion_free} by $u_n$ and take the difference:
\begin{equation}
   - T_{n-1,n} ( u_n f_{n-1} - u_{n-1} f_n      )  + T_{n,n+1} (u_{n+1}f_n - u_n f_{n+1}) = - f_n \sum_{m=0}^{N_\mathrm{pot}}V_{n,m} u_m. \label{ansn}
\end{equation} 
Here we have used $T_{n,n-1} = T_{n-1,n}$ because we have real-valued basis functions.
Similarly, multiply Eq.~\eqref{recursion_scat0} by $f_0$ and Eq.~\eqref{recursion_free0} by $u_0$, and take the difference:
\begin{equation}
    T_{0,1}(u_1 f_0 - f_1 u_0) = - f_0 \sum_{m=0}^{N_\mathrm{pot}}V_{0,m} u_m.
    \label{ansn0}
\end{equation}

Now take the sum of Eq.~(\ref{ansn}) from $n=1$ to some $N > N_\mathrm{pot}$, and use Eq.~(\ref{ansn0}) for the $n=0$ term of the sum.  
Most of the terms on the LHS cancel (the LHS of Eq.~(\ref{ansn0}) cancels the first term of 
the LHS of Eq.~(\ref{ansn}) for $n=1$, while the second terms of the LHS of Eq.~(\ref{ansn}) for $n=1$ 
cancels the first term of (\ref{ansn}) for $n=2$, and so on) leaving only:
\begin{equation}
    T_{N, N+1}(u_{N+1} f_N - u_N f_{N+1}) = - \sum_{n=0}^{N_\mathrm{pot}} f_n \sum_{m=0}^{N_\mathrm{pot}}V_{n,m} u_m.
\end{equation}
The sum over $n$ on the RHS originally went to $N$, but by assumption the contributions for $n > N_\mathrm{pot}$ are zero. 
Thus this holds for any $N  > N_\mathrm{pot}$, where we have the `free' solution, $u_N = A f_N + B g_n$, so that 
\begin{equation}
    u_{N+1} f_{N} - u_{N} f_{N+1} = B (f_N g_{N+1} - f_{N+1} g_{N})
\end{equation}
leaving us with 
\begin{equation}
    B \left[T_{N,N+1} ( f_N g_{N+1} - f_{N+1} g_{N}) \right] = - \sum_{n=0}^{N_\mathrm{pot}} f_n \sum_{m=0}^{N_\mathrm{pot} }V_{n,m} u_m.
    \label{sindeltaJmat}
\end{equation}
This is the analog of Eq.~\eqref{integralphaseshift} in $J$-matrix formalism. 
The combination $ f_N g_{N+1} - f_{N+1} g_{N}$ is the Casoratian~\cite{bang2000p}, the discrete analog of the Wronskian.

In the same way, we tackle the function $g_n$.
Following the same derivation as Eq.~(\ref{sindeltaJmat}), but using Eq.~(\ref{eq:irregular_Jmatrix}), we get for any $N > N_\mathrm{pot}$
\begin{equation}
     - A \left[T_{N,N+1} ( f_N g_{N+1} - f_{N+1} g_{N}) \right] = - \sum_{n=0}^{N_\mathrm{pot}} g_n \sum_{m=0}^{N_\mathrm{pot} }V_{n,m} u_m - \alpha_0 u_0.
\end{equation}
Consequently, the phase shift in the $J$-matrix framework is
\begin{equation}
    \tan \delta = \frac{B}{A} = -\frac{  \sum_{n=0}^{N_\mathrm{pot}} f_n \sum_{m=0}^{N_\mathrm{pot} }V_{n,m} u_m } { \alpha_0 u_0 + \sum_{n=0}^{N_\mathrm{pot}} g_n \sum_{m=0}^{N_\mathrm{pot} }V_{n,m} u_m}. 
    \label{deltadiscHO}
\end{equation}
Here the inhomogeneity term $\alpha_0 u_0$ is much less troublesome than in 
coordinate space. 
{Because, as discussed next in Sec.~\ref{explain_g}, 
$\alpha_0 = W(f,g)/f_0$, where $W(f,g)$ is the Wronskian of $f$ and $g$, one can immediately see the resemblance to Eq.~(\ref{deltacont}). Previous results such as Eq.~(7) in \cite{PhysRevC.81.034002}, also have a similar form. 
Our derivation is more 
general.}

Because the sums only go up to  $N_\mathrm{pot}$, we no longer need to compute the $u_n$ up to some $N_\mathrm{max} > N_\mathrm{pot}$. Below we demonstrate that in many cases one can use a surprisingly small value of $N_\mathrm{pot}$.  The only remaining issue is generating the full solutions $u_n$. While one can diagonalize the Hamiltonian matrix, that is, solving the matrix eigenvalue equation $\hat{H} \mathrm{u} = E \mathrm{u}$, for our examples below we instead solved the equivalent Lippmann-Schwinger (LS) equation,
\begin{eqnarray}
    | u \rangle = | f \rangle + 
    \frac{1}{E - \hat{H}} \hat{V} | f \rangle,
    \label{LSE}
\end{eqnarray}
where $|f\rangle$ is the free-space regular solution with energy $E$. This allows for fine control on discretization of the scattering energy $E$.
{
We note that one can compute the resolvent $(E-\hat{H})^{-1}$ efficiently 
via the ever-useful Lanczos algorithm~\cite{haydock1974inverse} and that this has long been successfully implemented 
in many-body methods such as the configuration-interaction nuclear shell-model~\cite{PhysRevLett.63.949,PhysRevC.46.R2153}.


 {Systems with multiple channels have been addressed using 
 overlaps in coordinate space~\cite{PhysRevA.83.022705} and in the Appendix of \cite{PhysRevC.86.044330}.  One generalizes  Eq.~(\ref{eq:l2asympote}) to couple 
 the asymptotic wave functions in each channel via the scattering $S$-matrix or the reaction matrix $K$. 
Solution of the resulting linear equations for $K$ or $S$ produces a generalization of Eq.~(\ref{deltadiscHO}), where the analogs of $A$ and $B$ are matrices; when there are $N_\mathrm{ch}$ open channels, one needs $N_\mathrm{ch}$ linearly independent solutions at energy $E$ or order to make the $A$ and $B$ analogs into invertible matrices.  Multiple channels have also been previously addressed in the $J$-matrix framework, albeit not by overlap relations~\cite{broad1976j,PhysRevA.76.062706}. 
Because of the complexity of these equations, and important additional issues such as inelastic channels and channels below threshhold, we leave it to future work.
}
 

Finally, although our presentation is in terms of neutral particle scattering, our formalism is  generalizable to  
scattering of charged particles. One  needs to choose a 
suitable basis (the Laguerre basis) and use the $J$-matrix representations of 
the free regular and irregular Coulomb functions~\cite{yamani1975j,bang2000p}.  
This brings additional technical issues: nuclear physics predominantly prefers the harmonic 
oscillator basis, and the ``analytic'' forms of the $f_n$ and 
$g_n$ require extraction from the real and imaginary parts 
of a complex-valued hypergeometric function $_2F_1$~\cite{yamani1975j}; as an alternative, 
treating  Coulomb as a long-range correction in a harmonic 
oscillator basis has also been explored~\cite{bang2000p,yanikov2025harmonic}.}



{
\subsection{The irregular function in an $L^2$-integrable basis}

\label{explain_g}

Here we explain Eq.~(\ref{limgtilde}) in more detail, that is, how one represents the irregular 
scattering solution $g$ in an $L^2$-integrable basis, and how it can 
be interpreted as the solution to an inhomogeneous differential equation. 
This development is based upon \cite{yamani1975j}, especially their Eq.~(20).

Both the $f_n$ and the $g_n$ satisfy Eq.~(\ref{eq:tridiagonal_KE}) for $n > 0$, but, as discussed 
above, only $f_n$ satisfies Eq.~(\ref{eq:regular_Jmatrix}) 
while $g_n$ must satisfy Eq.~(\ref{eq:irregular_Jmatrix}). 
Going back to coordinate space, this implies that $(\hat{T} -E) f(r) = 0$ but
\begin{equation}
(\hat{T}-E) \tilde{g}(r) = \alpha_0 \phi_0(r). \label{inhom_diff}
\end{equation}
Since $\phi_0(r) \rightarrow 0$  as $r\rightarrow \infty$, we regain asymptotically 
$(\hat{T}-E)\tilde{g}(r) = 0$.

Because Eq.~(\ref{inhom_diff}) is an inhomogeneous differential equation, one  can solve Eq.(\ref{inhom_diff}) using a Green's function,
\begin{equation}
    G(r,r^\prime) = \frac{g(r_>) f(r_<)}{W(f,g)}
\end{equation}
where $r_<$ is the lesser of $r,r^\prime$, and $r_>$ is the 
greater, and $W(f,g)$ is the Wronskian. The solution of Eq.~(\ref{inhom_diff}) is (leaving off the homogeneous solution $f(r)$)
\begin{eqnarray}
        \tilde{g}(r) & =& \int_0^\infty G(r,r^\prime) \, \alpha_0 \phi_0(r^\prime) \, dr^\prime \\
        & = & \frac{\alpha_0}{W} \left ( 
g(r)\int_0^r f(r^\prime) \phi_0(r^\prime) \, dr^\prime
+ f(r)\int_r^\infty g(r^\prime) \phi_0(r^\prime) \, dr^\prime
        \right). \label{inhom_soln}
\end{eqnarray}
The second term of Eq.~(\ref{inhom_soln}) dominates for $r$ small, 
and is regular as $r \rightarrow 0$, but as $r \rightarrow \infty$,
the first term dominates and is proportional to the irregular 
free solution. But because we want it to approach the irregular free solution as $r\rightarrow \infty$, we must have
\begin{equation}
    \frac{\alpha_0}{W} \int_0^\infty f(r^\prime) \phi_0(r^\prime) \, dr^\prime =  \frac{\alpha_0}{W}f_0 = 1
\end{equation}
or
\begin{equation}
    \alpha_0 = \frac{W}{f_0}.
\end{equation}
Thus we have shown that the representation of the irregular function in $L^2$-integrable functions, $\tilde{g}$, is a solution to an inhomogeneous differential equation. 

Finally, we note that several prior applications of overlap methods 
in coordinate space, e.g.~\cite{PhysRevC.81.034002,PhysRevC.86.044330}, 
use a regularized $g(r)$ to avoid difficulties of wave function evaluation at or near $r=0$}, multiplying the standard function by an empirical expression 
$(1-e^{-\gamma r})^{2l+1}$. As this expression $\rightarrow 1$ for large $r$ and $\rightarrow \sim r^{21 + 1}$ for small $r$, the product behaves like $g(r)$ at 
large $r$, but at small $r$ behaves like $f(r)$, and hence 
acts as an \textit{ad hoc} version of $\tilde{g}(r)$. Ref.~\cite{viviani2020fancyregularizer} develops a more elaborate regularizer that keeps both the modified $g(r)$ and the modified $(\hat{H}-E)g(r)$ well-behaved at the origin.  In both  continuous and discrete schemes, however,  the SOR only requires that the two free solutions have a known and nonzero Wronskian in the $V=0$ part of the model space.

In the next section, we illustrate how SOR works in the scattering problem by two simple potentials, as well as one realistic one.

\section{Example Calculation and Discussion}
\label{example}
In this section, we illustrate SOR at work in an $L^2$-integrable basis in computing the scattering phase shift. 
Although our ultimate goal is to apply SOR in many-body calculations,  we 
restrict ourselves here to one-body scattering from a potential, so that we can compare 
results against standard phase shift calculations.
In all the calculations below we set our basis size $N_\mathrm{max}=N_\mathrm{pot}$.

\begin{figure}[h]
    \includegraphics[width=0.9\textwidth]{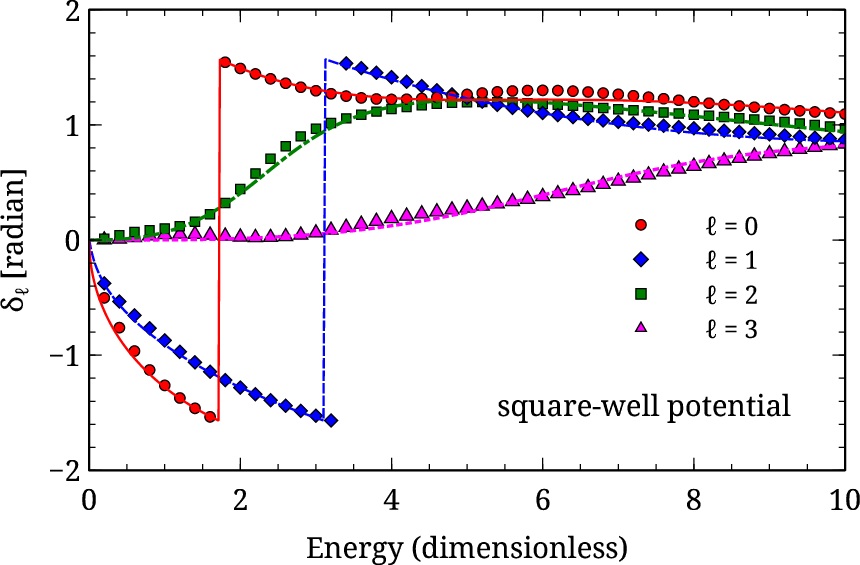}
    \caption{The scattering phase shift from the square-well potential as a function of (dimensionless) energy. The symbols are the scattering overlap relation (SOR) results for orbital angular momentum $\ell$ from 0 to 3. We solved the Lippmann-Schwinger equation with $N_{\rm{pot}} = 10$, that is, 11 oscillator basis states, and oscillator length $b = 0.42$, which minimizes the ground state energy in this truncated space.
    The lines are analytic solutions. The well has dimensionless radius $R=1$ and depth $V_0 = -5$; in addition, $\hbar=\mu=1$. { The discontinuities 
    are artifacts of keeping the phase shifts between $+\pi$ 
    and $-\pi$ radians.}}
    \label{fig:SWpot}
\end{figure}


We work in a harmonic oscillator (HO) basis, that is, our $\phi_n(r)$ are 3-D harmonic oscillator radial wave functions. In 
coordinate space the regular and irregular free solutions are 
spherical Bessel functions $f(r) = kr \,j_\ell(kr)$ and spherical Neumann function 
$g(r) = kr\,  y_\ell(kr)$, respectively; then 
 the $f_n$ and $g_n$ have analytic expressions (see \cite{PhysRevC.94.064320} and references therein):
\begin{eqnarray}
f_n & = & {\cal N}(n,\ell) q^{\ell+1} \exp (- q^2/2) L^{\ell+1/2}_{n}(q^2),
\\
g_n & = & {\cal N}(n,\ell) \frac{-q^\ell}{\Gamma(-\ell + 1/2)} \exp(-q^2/2) \, \\ \nonumber & & \times \Phi (-n-\ell-1/2;-\ell + 1/2; q^2),
\end{eqnarray}
where the overall normalization factor is given by
\begin{equation}
{\cal N}(n,\ell) =\sqrt{\frac{\pi \Gamma(n+1)}{\Gamma(n+\ell + 3/2)}}, 
\end{equation}
where $q = \sqrt{\frac{2E}{\hbar \omega}} = k b$, $L_n^m$ is the associated Laguerre function, and $\Phi$ is the confluent hypergeometric function.  (See \cite{muller2001computing,pearson2017numerical} for discussions 
of computing the confluent hypergeometric function at large values of the arguments.)
Here $n$ is the number of nodes in the radial wave function, related to the principal quantum number $N_\mathrm{princ}$  by $N_\mathrm{princ} = 2n + l$. 

Finally, we found the scattering solutions $u_n$  
by solving the LS equation (\ref{LSE}); although not
shown, separate calculations of the $u_n$ found by matrix diagonalization gave very good agreement.
The LS equation, however, allowed for finer sampling of the continuum because its energies are not restricted to a set of discrete eigenvalues.

Our first example is scattering off a square well potential, for which analytic solutions are easily found. For 
Fig.~\ref{fig:SWpot} we chose dimensionless parameters $\hbar = \mu = 1$, potential well radius $R = 1$, and well depth $V_0 = -5$.
We choose $N_\mathrm{pot} = 10$ and used a harmonic oscillator length parameter, $b = b_\mathrm{min} = \sqrt{\hbar/\mu \omega} =0.42$, where $\omega$ is the oscillator frequency. This choice of $b$ minimizes the ground state energy in this truncated space; in the Appendix, we discuss the relation between the ground state energy and the quality of the phase shifts. In a large harmonic oscillator basis ($N_\mathrm{pot} = 200$) the dimensionless ground state energy is $-2.312$,
with very little sensitivity to $b$, in agreement within numerical error with the result in coordinate space.  
Even in our smaller space, $N_\mathrm{pot} =10$, we get excellent agreement with the analytic phase shifts for orbital 
angular momentum $\ell = 0, 1,2,3.$

\begin{figure}
    \centering
    \includegraphics[width=0.9\textwidth]{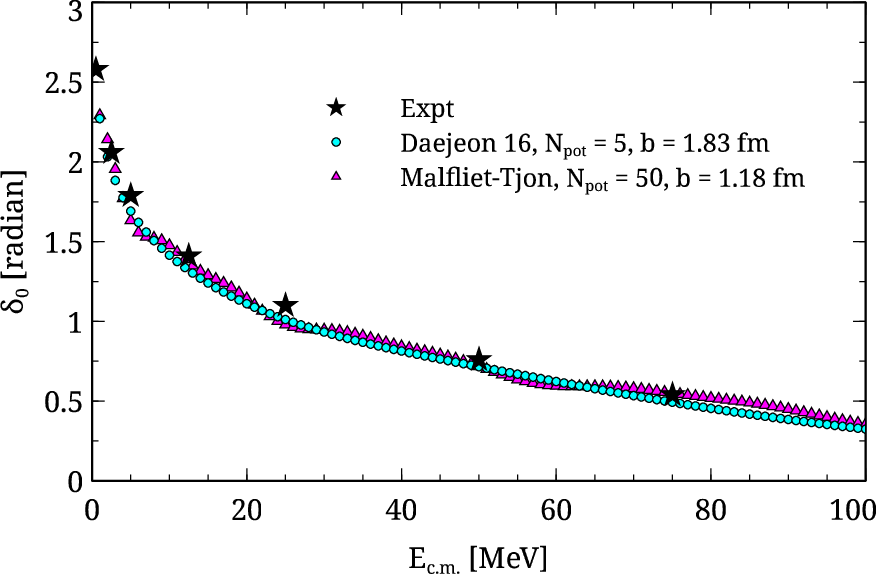}
    \caption{The isoscalar $s$-wave phase shifts using the Daejeon 16 potential (circles) with $N_{\rm{pot}} = 5$, that is, 6 oscillator basis states, and $b = 1.83$ fm (which corresponds to $\hbar\omega = 25$ MeV), and the Malfliet-Tjon potential (triangles) with $N_\mathrm{pot} = 50$ ($b = b_{\rm{min}}$ = 1.18 fm, $E_{\rm{g.s.}} = -2.287$ MeV). Experimental values~\cite{PhysRevC.48.792} are stars. We excluded coupling to $^{3}D_{1}$ partial wave in the calculation with Daejeon-16, resulting in a   ground state energy of -1.022 MeV. { The slight oscillations in the Malfliet-Tjon results do not appear if one solves in coordinate space, or in harmonic oscillator space with a larger $N_\mathrm{pot}$.}}
    \label{fig:Fig2}
\end{figure}

For a more realistic, and more challenging, case, 
we computed isoscalar $^3S_1$ nucleon-nucleon phase shifts. In
Fig.~\ref{fig:Fig2}, we compare against experimental 
phase shifts~\cite{PhysRevC.48.792} results using two interaction  
models.  The first is the   Malfliet-Tjon  potential~\cite{malfliet1970three}:
\begin{equation}
    V_{\text{MT}}(r) = V_a \frac{e^{-\mu_a r}}{r} + V_r \frac{e^{-\mu_r r}}{r},
\end{equation}
with parameter values $V_a = -635$ MeV, $V_r = 1458$ MeV, $\mu_a = 1.55$ fm$^{-1}$, and $\mu_r = 3.11$ fm$^{-1}$. 
These parameters yield a 
deuteron binding energy of $-2.29$ MeV, in rough agreement with  experimental value  $-2.22$ MeV \cite{Greene1986}. 
The other is the more modern Daejeon-16 
interaction~\cite{shirokov2016n3lo},  generated from chiral effective field theory interaction matrix elements~\cite{PhysRevC.68.041001} and softened through the similarity renormalization group and phase-equivalent transformations. Daejeon-16  exists only as matrix elements in harmonic oscillator space.

We computed the SOR phase shifts from Daejeon-16 in a small space, $N_\mathrm{pot}= 5$, using a harmonic 
oscillator length parameter $b = 1.83$ fm, which corresponds to a frequency of $\hbar \omega = 25$ MeV. 
Because we leave the coupled-channel formalism to future work, 
we excluded the coupling to the $^3D_1$ partial wave, which led to a poor value of the ground state energy, $-1.022$ MeV (if we include the $^3D_1$ we regain a good value of the ground state energy), but nonetheless get good agreement with experimental phase shifts.  We believe the soft nature of Daejeon-16 is important, because the Malfliet-Tjon interaction, 
with a strong repulsive core, proved more challenging. 
In Fig.~\ref{fig:Fig2}, using $N_\mathrm{pot}=50$ and a $b_\mathrm{min}=1.18$ fm, we obtained the general trend of the experimental phase shifts, albeit with some slight oscillations, which are not seen for phase shifts computed in coordinate space, { e.g., generated with the Numerov algorithm}. If we use a smaller $N_\mathrm{pot}$, however, the oscillations seen in Fig.~\ref{fig:Fig2}, grow in amplitude; {  if we use a larger $N_\mathrm{pot}$, the oscillations decrease.} 
{ It is plausible that with a different basis, Malfliet-Tjon would 
behave better for smaller $N_\mathrm{pot}$, but the $J$-matrix method assumes 
a tridiagonal matrix for the kinetic energy, Eq.~(\ref{eq:tridiagonal_KE}), 
for which the harmonic oscillator and the Laguerre bases have analytic expressions.  One might numerically generate a new basis using the 
Lanczos algorithm, which produces a tridiagonal form, an approach we are exploring. Alternately, there are generalized $J$-matrix approaches that do not 
require a tridiagonal form~\cite{PhysRevA.64.042703}. Clearly there remains 
much rich territory to explore.
}

\section{Conclusions}

In formalisms that use an explicit state vector, the scattering problem has two logical parts: generation of a scattering state and extraction from that state of phase shift and mixing parameters (i.e., the $S$-matrix).
 Focusing in this paper on the second part,
we have derived a generalized scattering overlap relation (SOR) for  single-channel phase shifts that applies both to continuous position basis 
calculations and to discrete, $L^2$-integrable bases such as the harmonic oscillator basis used in $J$-matrix calculations.  We then demonstrated successful application of the discrete SOR to various potentials, including at nonzero angular momentum.
Our general analysis illuminates the 
parallels between using overlap relations in coordinate 
space and in an $L^2$ basis.
In some cases we reproduce the phase shifts in a relatively small model space, in particular for a soft interaction such as Daejeon-16.
We found conversely that the Malfliet-Tjon potential, which has a strong short-range repulsive core, required a larger model oscillator space; it is not clear if all basis sets would require similarly large model spaces.
In the Appendix, we discuss the relation between reproducing phase shifts and reproducing the ground state energy.

{
This approach should be generalizable. While the bulk of 
applications of  the $J$-matrix method in nuclear physics, at least, have been for computing single-channel phase shifts, 
e.g.~\cite{PhysRevC.94.064320,PhysRevC.98.044624,mazur2019description,shirokov2019description,PhysRevC.106.064320},
multichannel scattering has been computed both in the $J$-matrix method~\cite{broad1976j,PhysRevA.76.062706} and in coordinate-space overlap integrals~\cite{PhysRevA.83.022705}; therefore combining them 
to compute $S$-matrices using scattering overlap relations in 
$L^2$-integrable bases is possible  by  a matrix generalization of our basic result in Eq.~(\ref{deltadiscHO}). In addition one can generalize to Coulomb scattering~\cite{yamani1975j,bang2000p} and 
to asymptotic normalization coefficients~\cite{PhysRevC.83.041001}.
{

With the use of SOR, it should be straightforward to address both state-generation and observable-extraction parts of the scattering problem in a many-body method such as configuration-interaction.  Here we simply sketch the basic ideas. Suppose we  consider the scattering of a particle 
such as a nucleon or an electron off an $A$-particle target, with the 
bound-state many-body wave function $ | A \rangle $. Let $\hat{\phi}_n^\dagger$ 
be the creation operator for a particle in the basis state $ \phi_n$; 
then we can add either a regular partial wave at a given momentum, 
\begin{equation}
    | f \rangle = \left ( \sum_n f_n \hat{\phi}_n \right ) | A \rangle,
\end{equation}
or an irregular partial wave,
\begin{equation}
    | g \rangle = \left ( \sum_n g_n \hat{\phi}_n  \right ) | A \rangle.
\end{equation}
(If one is concerned about spurious center-of-mass motion, one 
can easily project out eigenstates of the center-of-mass operator 
by a modified Lanczos algorithm~\cite{PhysRevC.91.034313,zbikowski2021rotational}.)
Finally, one can generate the fully interacting solution $| u \rangle$
by the Lippmann-Schwinger Eq.~(\ref{LSE}), for example applying the 
resolvant again by the Lanczos 
algorithm~\cite{haydock1974inverse,PhysRevLett.63.949,PhysRevC.46.R2153};
for coupled-channel $S$-matrices, at a given energy one would need to find as many independent solutions as there are open channels,}  and then operate on a matrix of amplitudes computed like the numerator and denominator of our basic result in Eq.~(\ref{deltadiscHO}).
This current work thus represents an intermediate but important step towards more realistic applications.

\section*{Acknowledgements}

We thank K.~Kravvaris (Lawrence Livermore National Laboratory) for helpful conversations and useful feedback, and YoungHo Song (IBS, Republic of Korea) for providing and discussing the Daejeon 16 potential matrix elements in the harmonic oscillator basis.
This material is based upon work supported by the U.S. Department of Energy, under Award Number DE-NA0004075.

\appendix

\section{Truncation, phase shifts, and the ground state energy }
\label{TruVar}
\begin{figure}
    \centering
    \includegraphics[width=0.9\textwidth]{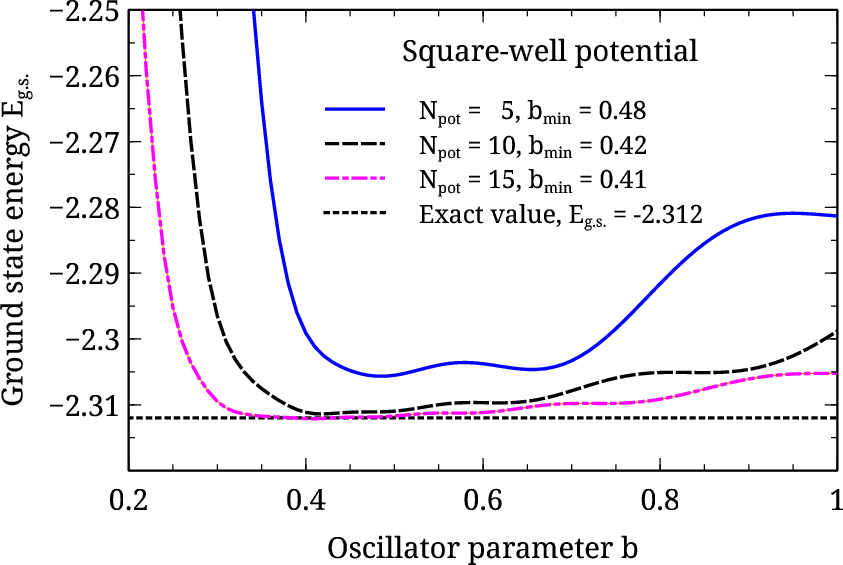}
    \caption{The ground state energy of the system when the parameter $b$ is varied. The exact value of the ground state energy (horizontal line) is $-2.312$ (dimensionless). $b_{\rm{min}}$ is the value that makes $E_{\text{g.s.}}$ minimum.
    }
    \label{fig:AppendixB1}
\end{figure}

Out of practicality, one must truncate the matrix representation 
of the potential $V_{n,n^\prime}$, such that $n,n^\prime \leq N_\mathrm{pot}$.
In tractable many-body calculations, the effective $N_\mathrm{pot}$ might 
be quite small: because of parity conservation, an $N_\mathrm{pot}$ of 5 is equivalent to twelve oscillator shells in a no-core nuclear shell model calculation~\cite{barrett2013ab}.  Nonetheless, our results demonstrate that one can reasonably reproduce the phase shifts for such a small 
$N_\mathrm{pot}$, especially for a soft interaction such as Daejeon-16.

In the course of our investigation, we 
found that the quality of the phase shifts correlated 
with the quality of the ground-state energy. (The cases we considered above each had exactly one bound state.) 
Specifically, for a fixed $N_\mathrm{pot}$, 
we varied the oscillator 
length parameter $b = \sqrt{\hbar/\mu \omega}$, where $\omega$ is the 
oscillator frequency.  We found  our `best' 
phase shifts when $b$ minimized the (bound) ground state energy. 

To demonstrate this,  consider the same square well potential as in Fig.~\ref{fig:SWpot},  which has a (dimensionless) ground state energy of $-2.312$ (that this is numerically similar to the deuteron's binding energy in MeV is coincidental). 
In Fig.~\ref{fig:AppendixB1} we compute the ground state energy, that is, 
the energy of the single $s$-wave bound state, in a basis of oscillator 
states, as a function of the oscillator length parameter $b$ for 
$N_\mathrm{pot}$ = 5, 10, and 15.

Unsurprisingly, calculations in larger spaces (larger $N_\mathrm{pot}$) are less sensitive to $b$ and are closer to the infinite space ground state energy.
(One can understand the sharp upturn at small $b$
  by appealing to an important property of the oscillator functions: the classical turning point,
\begin{equation}
    R_{cl} = b \sqrt{4N_\mathrm{pot} + 2 \ell +3}. \label{Rn}
\end{equation}
For $\ell = 0$ and for $N_\mathrm{pot} = 5$, we get approximately 
$R_{cl} \approx 4.8 b$, so when $b < 0.2$, the classical turning point is 
\textit{within} the well radius. In such a case it is unsurprising the ground state energy is badly estimated.)

\begin{figure}

\begin{tabular}{c}
\includegraphics[scale=0.8,clip]{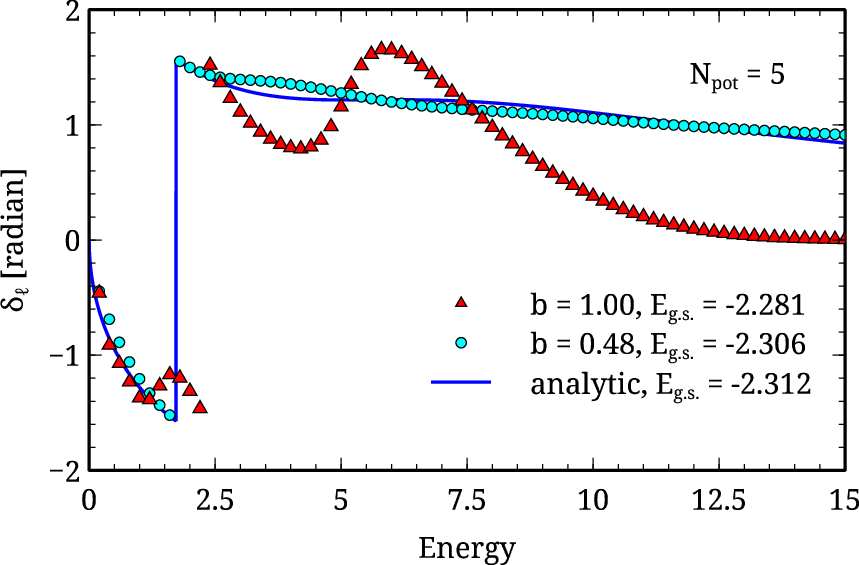} \\ 
\end{tabular}
    \caption{The scattering phase shift by the square-well potential with $N_\mathrm{pot} = 5$. For $b = 1.00$ and $0.48$, the bound state energies are $-2.281$ and $-2.306$ (dimensionless), respectively. The large-space ground state energy is -2.312. { The discontinuities 
    are artifacts of keeping the phase shift between $+\pi$ 
    and $-\pi$ radians.}
    }
    \label{fig:AppendixB2}
\end{figure}

Correlated with the ground state energy are the 
phase shifts.
 In Fig~\ref{fig:AppendixB2} we show the $s$-wave phase shift for $N_\mathrm{pot}=5$ and for two 
different values of $b$, compared with the exact analytic result.  Note that the phase shift is improved when the ground state energy is better approximated, a general result we found in our investigations. While this is not surprising, we have not found an effective argument to link bound state energies and phase shifts.
In many-body calculations in a harmonic oscillator basis~\cite{barrett2013ab}, one typically chooses the basis frequency (and thus the length parameter $b$), so as to minimize the ground state energy. This gives us hope that, even in a highly truncated basis, we can effectively recover scattering information. This will be part of future investigations.

\section{Technical issues near the origin}
\label{sec:tech_origin}

In this Appendix we discuss in some detail the $r\rightarrow 0$ limit of 
Eq.~(\ref{eq:surface_terms}), that is, 
\begin{equation}
    (g u^\prime  - u g^\prime)_{r \rightarrow 0}. \label{eq:limit_zero} 
\end{equation}
{For $\ell =0$, 
$g(0)$ is finite while $u(0)=0$, so that}
$-( g u^\prime  - u g^\prime )_{r = 0} = -g(0) u'(0)$. 
For $\ell > 0$, the irregular function $g(r) = kr \, y_\ell(kr)  \approx - (2\ell-1)!!\,  (kr)^{-\ell}$; and (as long as 
$\lim_{r \rightarrow 0} r^2 V(r) = 0$ so that the centrifugal term dominates as $r \rightarrow 0$),  for small $r$, we can write $u(r) = u_c \times r^{\ell+1} $. Then
$\lim_{r\rightarrow 0}g(r) u^\prime(r)- u(r) g^\prime(r)  =$ a finite constant. For simplicity, we  write 
this limit as $g(0) u^\prime(0) - u(0) g^\prime(0)$. We can evaluate this: 
\begin{eqnarray}
  \lim_{r \rightarrow 0} g(r) u^\prime(r) - u(r) g^\prime(r) \nonumber  \\
  =   \lim_{r \rightarrow 0}  -\frac{(2\ell-1)!!}{(kr)^\ell}
   (\ell +1) u_c r^\ell - u_c r^{\ell +1} (\ell ) \frac {(2 \ell-1)!!}{ k^{\ell} r^{\ell +1}} 
   \\ = - \frac{ (2\ell+1)!!}{k^{\ell}} u_c.
\end{eqnarray}
If we compute on a regular lattice of spacing $\Delta r$, 
we have $u_0 = u(0) = 0$ and $u_1 = u(\Delta r) = u_c  \times \Delta r^{\ell +1}$, and so
\begin{equation}
      \lim_{r \rightarrow 0} g(r) u^\prime(r) - u(r) g^\prime(r) = - \frac{ (2\ell+1)!!}{k^{\ell}}
      \frac{u_1}{(\Delta r)^{\ell+1}} = - k \frac{u_1}{ f_1},
\end{equation}
where we used $f_1 = f(\Delta r) $ and $f(r) = kr \, j_\ell(kr) = (k r)^{\ell+1}/(2\ell+1)!!$ for small $r$. Hence we can think of the limit $\lim_{r \rightarrow 0} g(r) u^\prime(r) - u(r) g^\prime(r)$ as the more tractable limit $\lim_{r \rightarrow 0} - k \, u(r)/f(r)$. 
Note that $k$ is the Wronskian of $f$ and $g$, a fact that will be echoed later. 
In some  applications, especially those where taking small $r$ is impractical, practitioners instead ``regularize'' the irregular function {so that 
$\lim_{r \rightarrow 0} g(r) u^\prime(r) - u(r) g^\prime(r) \rightarrow 0$}~\cite{PhysRevC.81.034002,PhysRevC.108.034001,viviani2020fancyregularizer}  while the usual Wronskian is retained at large $r$. As discussed in Section~\ref{discrete}, when using a square-integrable 
basis we do not face such difficulties because $g(r)$ is naturally regularized within the $J$-matrix approach as discussed in Section~\ref{explain_g}. The limit $g(0) u^\prime(0) - u(0) g^\prime(0)$ is an analog to the {inhomogeneity} introduced 
in the main text above, especially in Section \ref{explain_g}.

\bibliographystyle{unsrt}
\bibliography{jmatrix}

\end{document}